\documentclass[epj]{svjour}
\usepackage{graphics,amsmath,amssymb}
\begin{document}
\title{Entrance-channel potentials \\
in the synthesis of the heaviest nuclei}
%\subtitle{}
\author{V.Yu. Denisov \inst{1,2}
\thanks{\emph{e-mail addresses :} v.denisov@gsi.de, denisov@kinr.kiev.ua }
\and
W. N\"orenberg \inst{1,3}
\thanks{\emph{e-mail address :} w.nrnbrg@gsi.de}
} % Do not remove
%
%\offprints{} % Insert a name or remove this line
%
\institute{ Gesellschaft f\"ur Schwerionenforschung, D-64291
Darmstadt, Germany \and Institute for Nuclear Research, 03680 Kiev,
Ukraine \and Institut f\"ur Kernphysik, Technische Universit\"at
Darmstadt, D-64289 Darmstadt, Germany}
\date{Received: \today / Revised version: \today}
% The correct dates will be entered by Springer
%
\abstract{Entrance-channel potentials in nucleus-nucleus
collisions, relevant for the synthesis of superheavy elements, are
systematically studied within a semi-microscopic approach, where
microscopic nuclear densities of the colliding spherical or
deformed nuclei are used in semi-classical expressions of the
energy-density functional. From experimental data on fusion
windows evidence is found that the existence of pockets in the
entrance-channel potentials is crucial for fusion. Criteria for
the choice of best collision systems for the synthesis of
superheavy elements are discussed.
 % end of PACS codes
} %end of abstract
\maketitle
\section{Introduction and summary}
\label{intro}

The formation of compound nuclei by fusion of very heavy nuclei is
one of the outstanding problems of low-energy nuclear reactions.
Such processes play a key role in the production of superheavy
elements (SHEs) [1-19] and of heavy nuclei far from the
$\beta$-stability line, as for example $^{217}$U \cite{217U}.

Recently, various properties of SHEs have been studied
experimentally as well as theoretically [1-8,20-29]. However, the
dynamical processes leading to SHEs in heavy-ion fusion reactions
are not yet well understood. Few attempts have been made to
develop models, which yield cross-sections for SHE production
\cite{adamian,nasirov,cherepanov,kobe,smolfus,smolfus2,dh,d}.
These models are mainly concerned with the evolution from the
touching configuration of the colliding nuclei to the compound
nucleus. The dynamics has been treated as diffusion in mass
asymmetry at the touching point \cite{adamian,nasirov,cherepanov},
as collective motion in 3-dimensions within Langevin equations
\cite{kobe}, as a tunneling process of outer barrier
\cite{smolfus,smolfus2}, and as shape evolution and tunneling
\cite{dh,d}.

We suggest to consider the fusion process as part of the
quantum-mechanical collision process \cite{nd}, where the features
of the entrance channel are essential for the scattering solution.
In such a formulation the population of quasi-bound states near
touching is regarded as the first decisive step for fusion and is
similar to the doorway-state mechanism in resonance scattering
\cite{feshbach,weidenmuller}. The existence of such capture states
depends crucially on the detailed properties of the
entrance-channel potential as a function of the distance $R$
between the nuclei, in particular on the existence of a pocket
which serves as a source for the quasi-bound states. If the
capture pocket is deep and wide, many quasi-bound states exist.
Then the coupling to complex states is strong and the probability
of compound-nucleus formation is much larger than for shallow
pockets. Because of its crucial role on the initial stage of the
fusion process, a precise and systematic knowledge of the
interaction potentials between the colliding nuclei is needed.

For determining the interaction potentials between two nuclei
various methods have been introduced. Early attempts are due to
Bass \cite{bass74,bass}, who parametrized a simple analytical
expression for the potential, to Swiatecki and coworkers
\cite{prox77}, who introduced the proximity interaction of
leptodermous systems and to Krappe, Nix and Sierk, who used the
folding procedure with a phenomenological Yukawa-plus-exponential
interaction \cite{kns}. Recently, a folding procedure using the
Migdal force \cite{migdal} has been used
\cite{adamian,nasirov,cherepanov,antonenko} together with
phenomenological density distributions. Here like in
\cite{bass74,bass,prox77,kns} a simple point-point interaction
$Z_1 Z_2 e^2/R$ (charges $Z_1, Z_2$) has been used for the Coulomb
part at distances larger than the touching point. A quite large
variance of the interaction potentials near touching are
encountered, when these different methods are used. These
uncertainties allow for quite different conclusions on the fusion
mechanism in the synthesis of SHEs
\cite{adamian,nasirov,cherepanov,kobe,smolfus,smolfus2,dh}.
Therefore, the precise knowledge of the nucleus-nucleus potential
near the touching point may also improve our qualitative
understanding of the fusion process.

We evaluate the interaction potential by keeping the densities of
the colliding nuclei fixed. Due to the short collision time this
frozen-density approximation is appropriate for the definition of
the entrance-channel potential (cf. section 2). The interaction
energy between the nuclei is obtained from the energy density
functional according to a definition which has been introduced by
Brueckner et al. \cite{brueckner}. The Thomas-Fermi approximation,
extended to all second-order gradient terms (ETF), is used for the
evaluation of the kinetic-energy density functional
\cite{brack,brackbook}, while the nuclear interaction energy is
obtained from a Skyrme energy-density functional. The nucleon
densities of the colliding spherical or deformed nuclei are
calculated in the microscopic Hartree-Fock-Bogoliubov
approximation. Our method is semi-micro\-sco\-pic, because we use
the microscopic nucleon densities and the semiclassical ETF
approach and energy densities. We refer to the interaction
potential, which is obtained from this semi-microscopic approach,
as the semi-microscopic potential (SMP). Since the microscopic HFB
densities are used in the extended semi-classical energy-density
functionals for the kinetic part as well as for the nuclear (and
Coulomb) interaction parts, we consider these SMPs as accurate and
reliable references, in particular around the touching point.

Within our semi-microscopic approach we have studied the
entrance-channel potentials for various systems which are of
interest for the synthesis of SHEs (section 3). The systems have
been roughly divided into three groups for cold, hot and warm
fusion according to the lowest attainable excitation energy
($\approx 15,$ 25 or 35 MeV) of the compound system and the
corresponding number (1, 2 or 3) of neutrons which have to be
emitted for reaching the compound-nucleus ground-state.

As compared to light systems, the potential pockets are quite
shallow for the heavy systems, and hence host much less capture
states. The experimentally well studied fusion windows in the
lead-target systems reveal interesting correlations with detailed
properties of our semi-microscopic  entrance-channel potentials
(SMPs) for different projectiles (section 3.1). While the mean
energies of the fusion windows lie systematically 5 to 10 MeV
below the SMP barriers, their widths occur to be proportional to
the depths of the potential pockets, such that the fusion windows
become narrower with increasing charge of the projectiles. The
shift of the fusion window below the barrier indicates that
transfer or virtual-excitation (polarization) channels are
important for populating the quasi-bound capture states. The
correlation of the fusion-window width with the depth of the
pocket indicate the crucial role of the quasi-bound capture states
for the fusion process. Symmetric cold-fusion systems with almost
equal projectiles and targets exhibit almost no pockets, and hence
are less favorable for the synthesis of SHEs.

In the hot-fusion systems one uses prolate uranium and
transuranium nuclei as targets. Thus the entrance-channel
potential depends significantly on the orientation of the deformed
nucleus (section 3.2). From the study of lighter systems one has
concluded \cite{fushindr1,fushindr2} that fusion occurs only in
side collisions, where the projectile hits the targets on the
waist-line ($90^\circ$ with respect to its symmetry axis). If this
is assumed also for the experimentally studied fusion reactions
with $^{48}$Ca on prolate uranium and transuranium nuclei we find
the same location of the fusion window as in the lead-target
systems, i.e. 5 to 10 MeV below the SMP barrier. $^{40}$Ca
projectiles are less favorable, because they lead to higher
excitation energies, which reduces the evaporation-residue
cross-section. The systematic study of $^{238}$U and $^{252}$Cf
target systems show deeper pockets for $^{238}$U and even more for
$^{252}$Cf than for $^{208}$Pb  target systems, however at higher
compound-nucleus excitation energies. Only for very heavy systems
like $^{82}$Se+$^{238}$U=$^{320}126$ or
$^{68}$Ni+$^{252}$Cf=$^{320}126$ we encounter low excitation
energies like in cold-fusion systems.

The warm-fusion systems, formed with the oblate $^{198}$Pt target,
yield compound-nucleus excitation energies which lie in between
those with $^{208}$Pb and with $^{238}$U or $^{252}$Cf. Pockets
are obtained up to $^{100}$Mo+$^{198}$Pt=$^{298}$120, where the
compound-nucleus excitation energy is comparable to cold-fusion
systems (section 3.3).

In conclusion, there are indications that the existence of
quasi-bound capture states in the entrance-channel potential
pocket is crucial for the synthesis of SHEs. These capture states
are only populated in an energy window with a mean value 5 to 10
MeV below our SMP barrier and a width roughly given by the depth
of the pocket. Whenever deformed nuclei are involved, fusion is
expected through the most compact capture states (pockets).
Applying these considerations to the synthesis of SHE 118, we
conclude that there are two favorable systems: the hot-fusion
system (which actually is more like a warm-fusion system)
$^{48}$Ca+$^{252}$Cf at about 206 MeV center-of-mass energy and
the warm-fusion system (which actually is an almost cold-fusion
system) $^{96}$Zr+$^{198}$Pt at about 330 MeV center-of-mass
energy (section 4).

\section{Definition of a semi-microscopic potential (SMP) in the
entrance channel}

The ultimate aim is the description of fusion as part of the
quantum-mechanical scattering process between two nuclei. In such
a formulation one has to consider all kinds of channels which are
coupled (in general non-perturbatively) to the entrance channel of
the colliding nuclei. Thus one faces the problem of describing the
coupling of a large number of channels which include inelastic
excitations and transfer, quasi-fission, compound-nucleus
formation and fission. In order to cope with this complexity one
has to formulate the stationary scattering problem in a suitable
way, which is the subject of a forthcoming paper \cite{nd}. One of
the crucial steps in such a description is the coupling of the
entrance-channel wave function to more compact configurations. It
is obvious that the energy dependence of the entrance-channel
wave-function strongly affects the cross-sections for all
processes including those which involve compact-shape
configurations. Therefore, we are interested to understand as a
first step the basic properties of the entrance channel in fusion
reactions. We evaluate the interaction potential of the entrance
channel in a semi-microscopic approach keeping the densities of
the colliding nuclei fixed. In the following, details of
definitions and justifications are given for these
semi-microscopic potentials (SMPs).

\subsection{Entrance channel dynamics}

Whereas the entrance channel (elastic channel) is uniquely defined
outside the range of nucleus-nucleus interactions, there is
considerable freedom inside. In principle, any definition, which
asymptotically describes two nuclei in their ground states, is
possible if all relevant inelastic channels are included in the
solution of the scattering problem. However, from a physical point
of view a reasonable definition should account for the essential
collective dynamics in the region of overlap.

A crucial quantity, which characterizes the collective dynamics in
the entrance channel during the capture process, is the nuclear
interaction time $\tau_{\rm coll}$ (collision time) as compared to
the characteristic time $\tau_{\rm relax}$ for the relaxation of
the intrinsic nuclear state due to nucleon-nucleon interactions.
The collision time $\tau_{\rm coll}$ may be estimated by
\begin{eqnarray*}
\tau_{\rm coll} \approx \frac{\pi}{\omega_{\rm pocket}}= \pi
\left[\frac{m A_1 A_2}{(A_1+A_2) V^{\prime\prime}(R_{\rm pocket})}
\right]^{1/2} ,
\end{eqnarray*}
where $\omega_{\rm pocket}$ denotes the``oscillator frequency" of
the interaction potential pocket $V(r)$, $R_{\rm pocket}$ is the
center-to-center distance at the pocket, while $m$, $A_1$ and
$A_2$ are the bare nucleon mass and the number of nucleons in the
nuclei. A typical value for $\omega_{\rm pocket}$ in reactions
used for SHE production is given by $\hbar \omega_{\rm pocket}
\approx 4$ MeV. Therefore, typical collision times for such cases
are $\tau_{\rm coll}\approx 5 \cdot 10^{-22}$ s. This value has to
be compared to typical times for the relaxation of the intrinsic
nuclear state due to nucleon-nucleon interactions. This time is
estimated as \cite{bertsch} $$\tau_{\rm relax} \approx
\frac{\epsilon_F}{3.2 \sigma v_F \rho_0 \varepsilon^* } \approx
\frac{2 \cdot 10^{-22}}{\varepsilon^*/{\rm MeV}} {\rm s}, $$ where
$\epsilon_F$ and $v_F$ denote the Fermi energy and velocity,
respectively, $\sigma$ the averaged nucleon-nucleon cross-section,
$\rho_0$ the normal density of nuclear matter and $\varepsilon^*$
the excitation energy per nucleon. For reactions leading to SHEs
we have at touching $\varepsilon^* \lesssim  (5 \; {\rm MeV})/250
\approx 0.02$ MeV, and hence $\tau_{\rm relax}  \gtrsim 10^{-20}$
s, i.e. more than one order of magnitude larger than the collision
time $\tau_{\rm coll}$. Thus we conclude that the entrance channel
in the region of nucleus-nucleus overlap is well defined by the
fixed configuration of the colliding nuclei. With respect to the
nucleus-nucleus potential in the entrance channel this means that
the interaction energy of two overlapping nuclei with frozen
densities is relevant.

The frozen-density potential should however not be applied to very
large overlap. A suitable continuation into regions of compact
shapes would be the diabatic energies of the entrance-channel
configuration \cite{diab1,diab2}. Furthermore, this frozen-density
potential should be regarded as a suitable reference for a
distribution of diabatic potentials (and barriers) which are due
to the mixture of configurations in the approaching nuclei
\cite{diab2}.

\subsection{Frozen-density potential}

We consider the interaction potential $V(R, \Theta)$ between a
spherical nucleus and an axially symmetric nucleus as function of
the center-to-center distance $R$ between the nuclei and the angle
$\Theta$ between the symmetry axis of the deformed nucleus and the
line connecting the centers of the nuclei. Denoting the energies
of the interacting nuclei by $E_{12}(R, \Theta)$ and the energies
of the non-interacting nuclei by $E_1$, $E_2$, we define the
interaction potential by
\begin{eqnarray}
V(R,\Theta) = E_{12}(R,\Theta) - E_1 - E_2 .
\end{eqnarray}
In the frozen-density approximation these energies are determined
by the energy-density functional ${\cal E}[ \rho_{p}({\bf r}),
\rho_{n}({\bf r})]$, i.e. \cite{brueckner}
\begin{eqnarray}
\lefteqn{E_{12}(R,\Theta) = }\\ & & \int \; {\cal E}[
\rho_{1p}({\bf r})+\rho_{2p}(R,\Theta,{\bf r}), \rho_{1n}({\bf
r})+\rho_{2n}(R,\Theta,{\bf r})] \; d{\bf r}, \nonumber
\end{eqnarray}
\begin{eqnarray}
E_1 = \int \; {\cal E}[ \rho_{1p}({\bf r}), \rho_{1n}({\bf r})] \;
d{\bf r}, \\
E_2 = \int \; {\cal E}[ \rho_{2p}({\bf r}), \rho_{2n}({\bf r})] \;
d{\bf r},
\end{eqnarray}
where $\rho_{1p}$, $\rho_{2p}$, $\rho_{1n}$ and $\rho_{2n}$ are
the frozen proton and neutron densities of the spherical nucleus
(index 1) and the deformed nucleus (index 2), respectively.

\subsection{Energy-density functional}

For an accurate calculation of the interaction potential between
two nuclei we need an energy-density functional which well
describes both the bulk and surface properties of the nuclei.
Suitable semi-classical expressions have been obtained for Skyrme
interactions and by an extended Thomas-Fermi (ETF) approximation
to the intrinsic kinetic energies, which includes all terms up to
second order in the spatial derivatives.

According to \cite{brack} the following expression for the
energy-density functional
\begin{equation}
{\cal E}[ \rho_{p}({\bf r}), \rho_{n}({\bf r})] =
\frac{\hbar^2}{2m} [\tau_p({\bf r}) +\tau_n({\bf r})] + {\cal
V}_{\rm sk}({\bf r}) + {\cal V}_{\rm c}({\bf r})
\end{equation}
has been deduced. The kinetic parts for protons ($i=p$) and
neutrons ($i=n$) are given by
\begin{eqnarray} \tau_{i}({\bf r}) =
\frac{3}{5}(3\pi^2)^{2/3} \rho_{i}^{5/3} + \frac{1}{36} \frac{(
\nabla \rho_{i} )^2}{ \rho_{i} } +
\frac{1}{3} \Delta \rho_{i} \;\;\; \\
+\frac{1}{6} \frac{ \nabla \rho_{i} \nabla f_{i} + \rho_{i} \Delta f_{i} }{
f_{i} }
- \frac{1}{12} \rho_{i} \left( \frac{\nabla f_{i}}{f_{i}}
\right)^2 \nonumber \\
+ \frac{1}{2} \rho_{i} \; \left( \frac{2m}{\hbar^2} \; \frac{W_0}{2}
\; \frac{ \nabla (\rho + \rho_{i}) }{f_{i}} \right)^2 \nonumber ,
\end{eqnarray}
where $W_0$ denotes the strength of the Skyrme spin-orbit
interaction, while $\rho=\rho_p+\rho_n$ and
\begin{equation}
f_{i}({\bf r}) = 1 + \frac{2m}{\hbar^2} \left( \frac{3t_1+5t_2}{16}
+\frac{t_2 x_2}{4}\right) \rho_{i}({\bf r}).
\end{equation}
The nuclear interaction part ${\cal V}_{\rm sk}$ results from the
Skyrme force and reads
\begin{eqnarray}
{\cal V}_{\rm sk}({\bf r}) \; = \; \frac{t_0}{2} \;
[(1+\frac{1}{2}x_0) \rho^2 -
(x_0+\frac{1}{2}) (\rho_p^2+\rho_n^2)] \;\\
+\frac{1}{12} t_3 \rho^\alpha [(1+\frac{1}{2}x_3 )\rho^2 -
(x_3+\frac{1}{2}) (\rho_p^2+\rho_n^2) ] \nonumber \\
+\frac{1}{4} [t_1(1+\frac{1}{2}x_1)+t_2(1+\frac{1}{2}x_2)] \tau \rho
\nonumber \\
+\frac{1}{4} [t_2(x_2+\frac{1}{2}) - t_1(x_1+\frac{1}{2})] (\tau_p
\rho_p+\tau_n \rho_n)
\nonumber \\
+\frac{1}{16}[3t_1(1+\frac{1}{2} x_1)-t_2(1+\frac{1}{2}x_2)] (\nabla
\rho)^2 \nonumber \\
- \frac{1}{16}[3t_1(x_1+\frac{1}{2} )+t_2(x_2+\frac{1}{2})] (\nabla
\rho_n)^2 +(\nabla \rho_p)^2 ) \nonumber \\
- \frac{W_0^2}{4} \frac{2m}{\hbar^2} \left[\frac{\rho_p}{f_p}(2\nabla
\rho_p+\nabla \rho_n)^2+ \frac{\rho_n}{f_n}(2\nabla \rho_n+\nabla
\rho_p)^2 \right] , \nonumber
\end{eqnarray}
where $t_0$, $t_1$, $t_2$, $x_0$, $x_1$, $x_2$, $\alpha$ and $W_0$
are Skyrme-force parameters. The Coulomb-energy density is
determined by
\begin{eqnarray}
{\cal V}_{\rm c}({\bf r}) = \frac{e^2}{2} \rho_p ({\bf r}) \int \;
\frac{\rho_p ({\bf r}' )}{|{\bf r}-{\bf r}' |}
d {\bf r}' \\
-\frac{3e^2}{4} \left( \frac{3}{\pi} \right)^{1/3} (\rho_p({\bf
r}))^{4/3}, \nonumber
\end{eqnarray}
where the last term is the local approximation to the exchange
contribution.

Thus, if the proton and neutron density distributions in both
nuclei are known, the interaction potential can be calculated from
the semi-classical expressions (1)--(9).

\subsection{Determination of density distributions}

The charge densities of nuclei are well described in Hartree-Fock,
Hartree-Fock-Bogoliubov (HFB) and semi-classical approaches
\cite{feshbach,brack,shfb}. However, HFB describes best various
other ground-state properties of nuclei \cite{shfb,rs} and
therefore has been chosen here for calculating the density
distributions.

For spherical nuclei we have used the HFB code \cite{shfb} with
very small radial mesh-point intervals of 0.025 fm. The gradients
and laplacians of the corresponding densities are evaluated
numerically by using Lagrange formulas. The density distributions
of deformed nuclei are also obtained in the HFB approximation by
using the code HFODD (v. 1.75r) \cite{defhfb}. The densities,
gradients and laplacians at any point of space are found by
Lagrange interpolation based on output from the HFODD code. This
code calculates the proton and neutron densities and their
gradients and laplacians at special points which are related to
the Gauss-Hermite integration. For an accurate evaluation of the
densities and their derivatives we use 48 points for the
Gauss-Hermite integration for each of the three space directions.

Note that the semi-classical approach based on microscopic HFB
densities is quite accurate for the calculation of binding
energies. For example, the differences of the binding energies
between HFB and the semi-classical expression are only $-0.9$ MeV
for $^{48}$Ca and $-0.3$ MeV for $^{208}$Pb, when the Skyrme force
SkM$^*$ is used. Moreover, differences in the energies cancel to a
large extent in the expression (1), such that the accuracy is
expected to be even higher for the nucleus-nucleus potential SMP.

\subsection{Relation to similar calculations}

The frozen-density approximation has been frequently applied
\cite{adamian,nasirov,antonenko,dp} in determining the
nucleus-nucleus potentials for the synthesis of SHEs. In
contradistinction to our approach, however, there are several
differences. For example in \cite{adamian,nasirov,antonenko,dp}
the Coulomb interaction potential is approximated by that of two
point charges or of homogeneous charge distributions,
respectively. In \cite{adamian,nasirov,antonenko} the nuclear part
of the nucleus-nucleus potential is calculated from a folding
procedure with Landau-Migdal interactions \cite{migdal}. However,
since this effective interaction is tailored to describe the force
between quasiparticles in the Fermi-liquid, the saturation
properties of nuclear matter cannot be obtained. Furthermore, the
nuclear density distributions, used for the folding procedure in
\cite{antonenko,dp}, are chosen as Fermi distributions with fitted
values of the radius and the diffuseness parameters. The kinetic
energy density was limited in \cite{dp} to the Thomas-Fermi
contribution which is the first term in (7). Similar calculations
for light nuclei have been performed in \cite{gupta}.

\section{Entrance-channel potentials}

As described in section 2, the nucleus-nucleus potential is
evaluated by numerical integration of the semi-classical
energy-density functional with the (frozen) HFB nucleon densities
of the separated nuclei. The integrals in Eqs. (2)-(4) are
evaluated by using the Gauss-Legendre method with a suitable
number of dots. The numerical integration is performed in
cylindrical coordinates. The system of two spherical nuclei is
axially symmetric, and hence the nuclear part of the potential is
reduced to the calculation of 2-dimensional integrals, while the
Coulomb part is obtained numerically from 3-dimensional integrals.
Since axial symmetry is lost for the system of a spherical nucleus
and a deformed nucleus, the nuclear and Coulomb parts have to be
calculated from 3- and 5-dimensional integrals, respectively.

\begin{figure}
% Use the relevant command for your figure-insertion program
% to insert the figure file.
% For example, with the option graphics use
\resizebox{9.cm}{!}{%
 \includegraphics{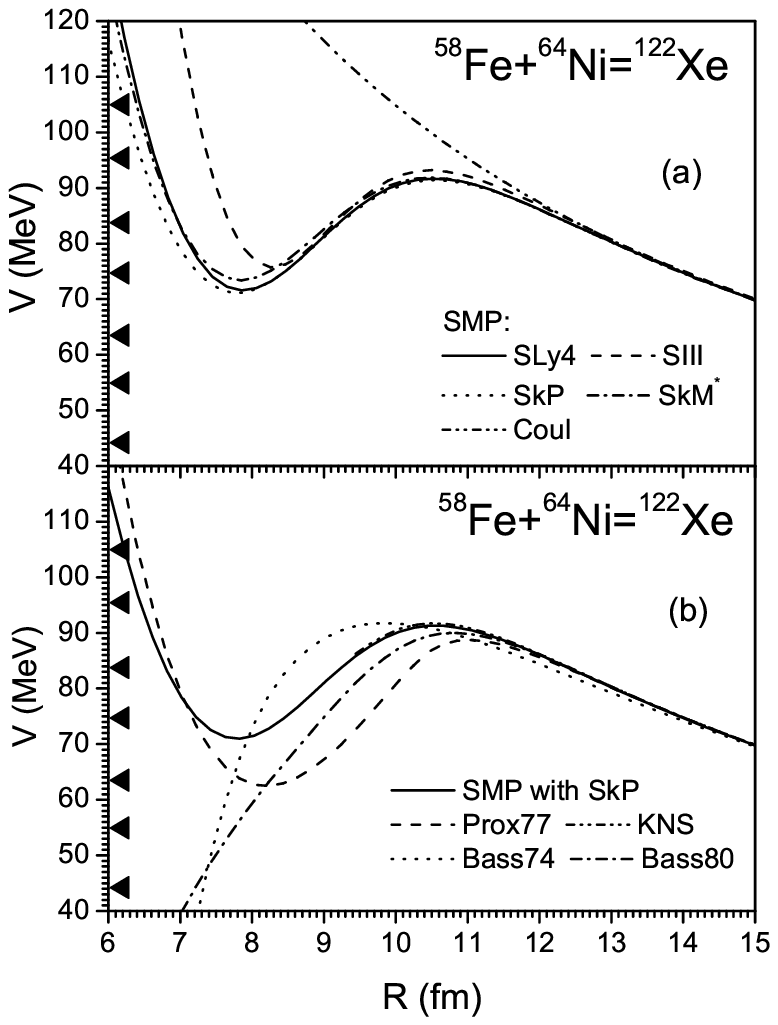}
}
% If not, use
%\vspace{0cm} % Give the correct figure height in cm
\caption{{\bf (a)} Semi-microscopic potentials (SMPs) for the
system $^{58}$Fe+$^{64}$Ni evaluated with the Skyrme forces SIII,
SkM$^*$, SkP and SLy4. For reference also the Coulomb potential is
presented. The ground-state Q-values are indicated by the lowest
triangle at the left vertical axis. The other 6 triangles mark,
respectively, the thresholds for the emission of 1, 2, 3, 4, 5 and
6 neutrons. {\bf (b)} SMP for the same collision system
$^{58}$Fe+$^{64}$Ni obtained with the Skyrme force SkP in relation
to the proximity potential (1977), the Bass potentials (1974,
1980) and the KNS potential.}
\label{fig:1} % Give a unique label
\end{figure}

In Fig. 1a we present the SMPs for $^{58}$Fe+$^{64}$Ni evaluated
for different Skyrme forces \cite{s3,skm,skp,sly4}. The potentials
obtained for SkM$^*$, SkP and SLy4 are very close to each other at
all distances down to $R=6$ fm. Considerable repulsion is observed
when the nuclei overlap and the density is doubling. Due to the
large value of the compression modulus (stiff equation of state),
this repulsion is particularly large for the SIII force and causes
the discrepancy with respect to the other Skyrme forces. In the
following sections 3.1 to 3.3 either SkP or SkM$^*$ is used, which
practically give the same interaction potentials.

The ground-state $Q$-value ($Q=E_{\rm CN}-E_1-E_2$) is obtained
from the experimental energies $E_1$, $E_2$ of projectile and
target taken from \cite{audi}. The compound-nucleus energy $E_{\rm
CN}$ is also obtained from this table or, if experimental values
are absent, from the Thomas-Fermi approach to nuclear masses
\cite{ms}. The neutron separation energies, which mark the
thresholds for neutron emission in Fig. 1, are also deduced from
\cite{audi,ms}.

Due to the deep pocket inside the barrier, light ions easily fuse
after tunneling through or passing over the barrier. Both, the
barrier height and the potential pocket are well above the
ground-state energy, such that the adiabatic potential surface
exhibits large gradients in the fusion direction driving the
system into the compound-nucleus shape.

The barriers obtained from different analytical expressions for
the nucleus-nucleus potential introduced by Bass in 1974
\cite{bass74} and in 1980 \cite{bass}, by Swiatecki et al.
(proximity 77) \cite{prox77} and by Krappe-Nix-Sierk (KNS)
\cite{kns} are spread over a large interval as shown in Fig. 1b.
The difference between the barriers is around 3 MeV in energy and
1 fm in position. The KNS barrier is closest to the SMP barrier.
Since the KNS potential depends on the shape parametrization for
distances smaller than touching, results for this potential is
presented only up to the touching point, which is given by $R =
r_0 (A_1^{1/3}+A_1^{1/3})$ with $r_0=1.2$ fm.

\subsection{Cold-fusion systems}

In this section we present results on entrance-channel potentials
of spherical projectiles and targets, which have been used in the
synthesis of SHEs by cold-fusion
\cite{hm,armbruster,munzenberg,hofmann}, and compare these
potentials with those of symmetric systems with almost equal
projectiles and targets.

\subsubsection{Cold-fusion systems with $^{208}$Pb targets}

\begin{figure*}
% Use the relevant command for your figure-insertion program
% to insert the figure file. See example above.
% If not, use
\begin{center}
\resizebox{14.4cm}{!}{%
 \includegraphics{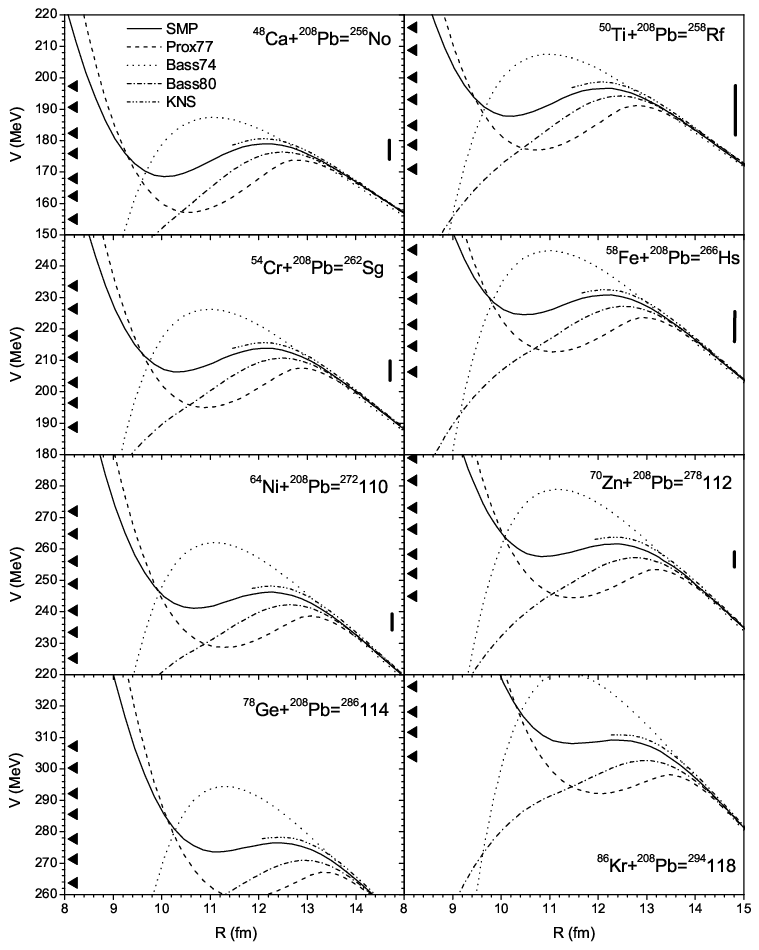}}
\end{center}
%\vspace*{16cm} % Give the correct figure height in cm
\caption{SMPs for the collision systems $^{48}$Ca+$^{208}$Pb,
$^{50}$Ti+$^{208}$Pb, $^{54}$Cr+$^{208}$Pb, $^{58}$Fe+$^{208}$Pb,
$^{64}$Ni+$^{208}$Pb, $^{70}$Zr+$^{208}$Pb, $^{76}$Ge+$^{208}$Pb,
and $^{86}$Kr+$^{208}$Pb obtained with the SkP Skyrme force as
compared to the proximity potential (1977), the Bass potentials
(1974, 1980) and the KNS potential. The ground-state Q-values are
indicated by the lowest triangles at the left vertical axes. The
other 6 triangles mark, respectively, the thresholds for the
emission of 1, 2, 3, 4, 5 and 6 neutrons. The observed fusion
windows [1--4] are indicated by the vertical bars on the
right-hand sides of the individual diagrams. }
\label{fig:2} % Give a unique label
\end{figure*}

Cold-fusion reactions with light projectiles ($Z \approx 20...36$)
on lead or bismuth targets have been used in the synthesis of SHEs
\cite{hm,armbruster,munzenberg,hofmann}. In Fig. 2 the interaction
potentials are presented for the systems with lead as target and 8
projectiles from $^{48}$Ca till $^{86}$Kr.  The following features
are observed.
\begin{itemize}
\item[--] The interaction potentials, which are obtained from
different standard expressions \cite{bass,prox77,kns}, are spread
over even larger intervals for heavier systems as compared  to
$^{58}$Fe+$^{64}$Ni (cf. Fig. 1b). As for the light system the KNS
potential is closest to our SMP around the barrier followed by the
Bass potential from 1980 \cite{bass}.

\item[--] The potential pockets are much shallower than for
$^{58}$Fe+$^{64}$Ni and tend to vanish with increasing size of the
projectile. For $^{96}$Zr+$^{208}$Pb (not shown in Fig. 2) no
pocket exists anymore.

\item[--] Since we consider the depth of the pockets to be
important for the fusion probability, we attribute the observed
\cite{hm} reduction of SHE formation with increasing size of the
projectile, at least partially, to the decreasing pocket depth.

\item[--] The observed fusion windows (vertical thick bars in Fig.
2) lie systematically about 5 to 10 MeV below our barriers. This
subbarrier fusion leading to SHEs is probably related to the
distribution of barriers and pockets due to the structure of the
approaching nuclei \cite{diab2} and/or to transfer and capture in
transfer-channel pockets, and possibly to other degrees of
freedom.

\item[--] A correlation is indicated between the width of the
observed fusion window and the depth of potential pocket (see
cases $^{50}$Ti+$^{208}$Pb, $^{58}$Fe+$^{208}$Pb and
$^{64}$Ni+$^{208}$Pb in Fig. 2).

\item[--] The difference between the barrier position and the
ground-state $Q$-value for fusion decreases with increasing charge
of the projectile, cf. \cite{hofmann,prox2000}. Due to the small
differences between the barrier heights and the ground-state
energies large gradients towards the compound nucleus are missing
in the adiabatic potential. Therefore, the shape evolution from
the touching configuration to the compound nucleus becomes
extremely sensitive to details of the potential landscape in
cold-fusion reaction for SHEs with $Z \gtrsim  110$ \cite{dh,d}.
Note that the observed fusion windows in cold-fusion reactions lie
about (10 -- 15) MeV above the ground-state $Q$-value
\cite{hm,hofmann}.

\end{itemize}

\subsubsection{Symmetric systems}

\begin{figure*}
% Use the relevant command for your figure-insertion program
% to insert the figure file. See example above.
% If not, use
\begin{center}
\resizebox{13.5cm}{!}{%
 \includegraphics{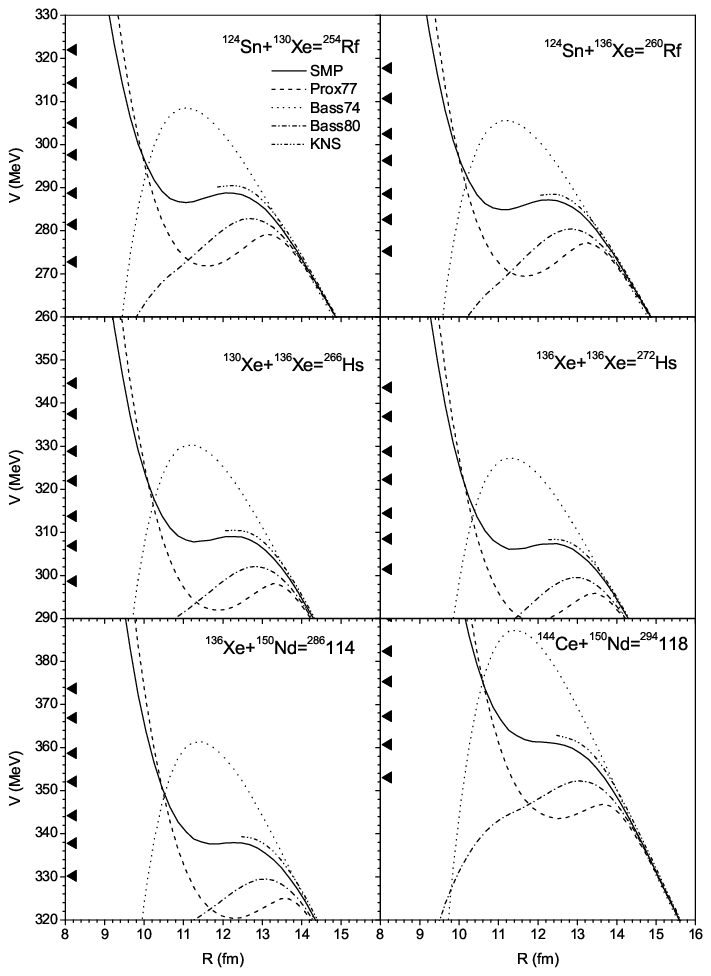}}
\end{center}
%\vspace*{14cm} % Give the correct figure height in cm
\caption{SMPs for the collision systems $^{124}$Sn+$^{130}$Xe,
$^{124}$Sn+$^{136}$Xe, $^{130}$Xe+$^{136}$Xe,
$^{136}$Xe+$^{136}$Xe, $^{136}$Xe+$^{150}$Nd and
$^{144}$Ce+$^{150}$Nd obtained with the SkP Skyrme force and the
proximity potentials (1977), the Bass potentials (1974, 1980) and
the KNS potential. The notations are the same as in Fig. 2. }
\label{fig:3} % Give a unique label
\end{figure*}

SHEs might also be formed by the fusion of similar nuclei, such as
Sn+Xe, Xe+Xe, Xe+Nd and others. The potentials for the collision
systems $^{124}$Sn+$^{130}$Xe, $^{124}$Sn +$^{136}$Xe,
$^{130}$Xe+$^{136}$Xe, $^{136}$Xe+$^{136}$Xe,
$^{136}$Xe+$^{150}$Nd and $^{144}$Ce+\\$^{150}$Nd are presented in
Fig. 3. As compared to the cold-fusion systems with $^{208}$Pb
targets (Fig. 2) the differences between the barrier heights and
the ground-state $Q$-values are smaller for symmetric systems
leading to the same SHE. Also the potential pockets are shallower
and at somewhat larger distances. Therefore, as compared to the
asymmetric systems of cold-fusion, the symmetric systems seem to
be less favorable because of the following reasons.

\begin{figure*}
% Use the relevant command for your figure-insertion program
% to insert the figure file.
% For example, with the option graphics use
\begin{center}
\resizebox{17cm}{!}{%
 \includegraphics{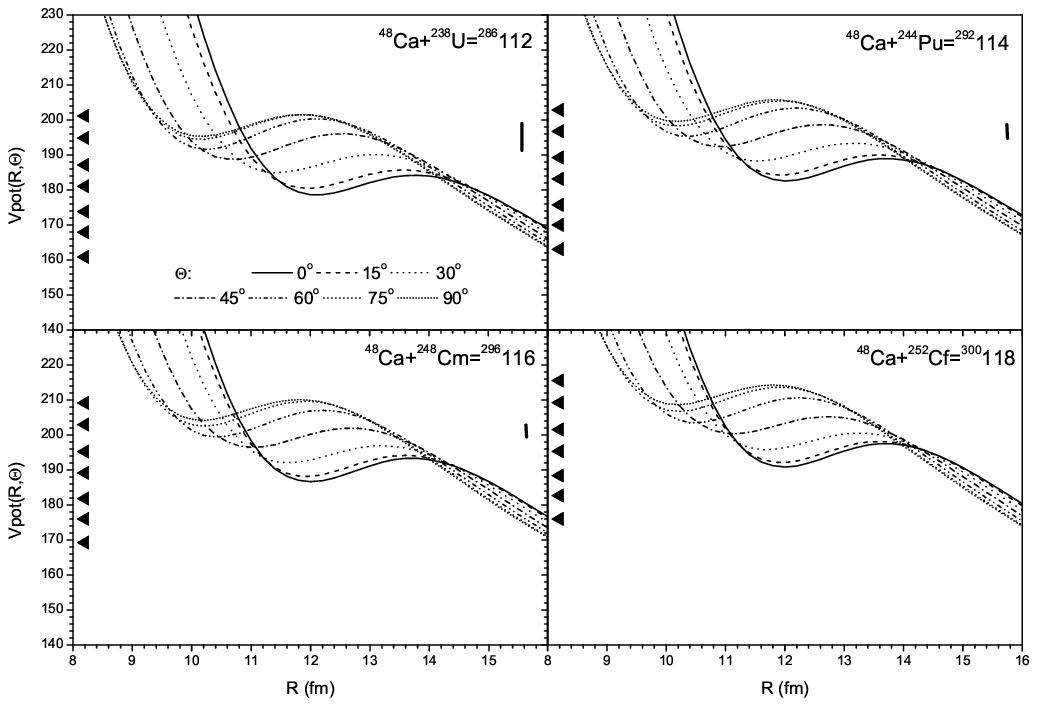}
}
% If not, use
%\vspace{0cm} % Give the correct figure height in cm
\end{center}
\caption{The SMPs for the collisions $^{48}$Ca on $^{238}$U,
$^{244}$Pu, $^{248}$Cm and $^{252}$Cf evaluated with the SkM$^*$
Skyrme force. The SMPs are evaluated for different angular
orientations of the heavy deformed nuclei. The ground-state
Q-values are indicated by the lowest triangle at the left vertical
axis. The other 6 triangles mark, respectively, the thresholds for
the emission of 1, 2, 3, 4, 5 and 6 neutrons. The bars at the
right-hand sides indicate the bombarding energies [6-8].}
\label{fig:4} % Give a unique label
\end{figure*}

\begin{itemize}
\item[--] The capture process is suppressed by the shallowness of
the potential pocket.

\item[--] The shape of the system at capture is less compact, and
hence a longer shape evolution is needed to reach the
compound-nucleus configuration, such that the formation
probability of the compound nucleus is reduced by the larger
competition of other decays.

\item[--] The capture windows are expected to lie 5 to 10 MeV
below the barriers, and hence are, apart from the lightest system
$^{124}$Sn+$^{130}$Xe, below the 1n fusion threshold.

\end{itemize}

\subsection{Hot-fusion systems}

\begin{figure*}
% Use the relevant command for your figure-insertion program
% to insert the figure file.
% For example, with the option graphics use
\begin{center}
\resizebox{17cm}{!}{%
 \includegraphics{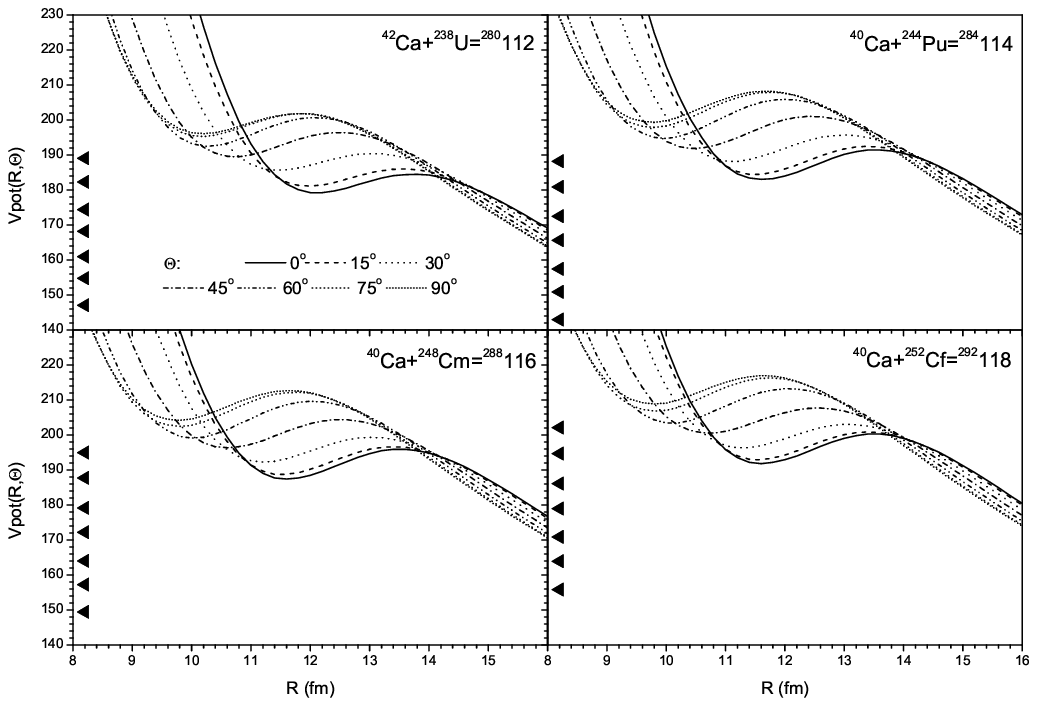}
}
% If not, use
%\vspace{0cm} % Give the correct figure height in cm
\end{center}
\caption{The SMPs for the collision $^{40,42}$Ca on $^{238}$U,
$^{244}$Pu, $^{248}$Cm and $^{252}$Cf evaluated with the SkM$^*$
Skyrme force. The SMPs are evaluated for different angular
orientations of the deformed nuclei. The notations are the same as
in Fig. 4.}
\label{fig:5} % Give a unique label
\end{figure*}

The fusion of light nuclei with heavy deformed nuclei (U, Pu, Cm)
leads already for collisions at barrier energies to large
excitation energies of about (30 -- 50) MeV in the compound
nucleus. Therefore, in such hot-fusion reactions typically 3 to 5
neutrons have to be emitted in order to reach the compound-nucleus
ground state. The projectiles are spherical and the targets are
well-deformed prolate nuclei in these hot fusion reactions.

\subsubsection{Ca projectiles}

Successful syntheses of SHEs with $Z=112$, 114 and 116 have been
reported \cite{112,114,116} for $^{48}$Ca projectiles on the
targets $^{238}$U, $^{244}$Pu and $^{248}$Cm. In this section we
discuss the entrance-channel potentials for the systems
$^{48}$Ca+$^{238}$U, $^{244}$Pu, $^{248}$Cm and $^{252}$Cf and
compare these with the corresponding potentials for $^{40,42}$Ca.

The interaction potentials as obtained from our semi-microscopic
method are shown in Fig. 4 for various orientations of the
deformed nuclei. For these systems the lowest barriers are
obtained for $\Theta =0^\circ$, i.e. when $^{48}$Ca touches the
tip of the deformed nucleus, while the barrier is highest for
$\Theta =90^\circ$, when $^{48}$Ca touches the side (cf.
\cite{fushindr1,fushindr2,denisov2,im} for lighter systems).

Outside the range of nuclear interactions the Coulomb interaction
tends to rotate the deformed nucleus into the $\Theta =90^\circ$
position (side position). However, the time for such a rotation is
typically
$$\tau_{\rm rot}\approx \frac{\pi/2}{\omega_{\rm rot}} = 2 \cdot 10^{-20}
\; {\rm s} \; ,$$ where a rotational energy $\hbar \omega_{\rm
rot}\approx 50$ keV has been inserted. Characteristic collision
times on the approaching part of the Coulomb trajectory are of the
order $2 \cdot 10^{-21} $ s \cite{coultime}, and hence the angle
of rotation during the approach is negligible.

As seen in Fig. 4 the positions of barriers and pockets depend
quite strongly on the orientation of the deformed nucleus, the
shift in distance and energy reaching 2 fm and 20 MeV,
respectively, between tip and side positions. In general, the
difference between the barrier and the ground-state energy of the
compound system is considerably higher in these very asymmetric
(hot-fusion) systems as compared to the cold-fusion systems of
section 3.1. Therefore, SHEs can be formed only in 3n to 4n
reactions, which reduces the survival probability strongly due to
the small branching ratio $\Gamma_n/\Gamma_f$ of neutron emission
to fission.

The bombarding energies in the fusion reactions $^{48}$Ca on
$^{238}$U, $^{244}$Pu, $^{248}$Cm \cite{112,114,116} have been
determined from the maximum cross-section for symmetric fission
events which indicate the formation of a compact relatively
long-living compound nucleus \cite{112,itkis}. These energies are
indicated by vertical bars in Fig. 4. For the cold-fusion
reactions with $^{208}$Pb targets (cf. Fig. 2) the fusion windows
lie about 5 to 10 MeV below our SMP barriers. If we assume that
the fusion mechanisms are similar also for the deformed targets --
and we do not see any reason to doubt this -- we have to conclude
from Fig. 4 that the side orientation ($\Theta \approx 90^\circ$)
is the relevant fusion channel for the formation of the SHEs. This
conclusion is supported by the experimental analysis of fusion
reaction between lighter nuclei \cite{fushindr1,fushindr2}, which
show that fusion through the tip orientation ($\Theta = 0^\circ$)
is strongly suppressed. Moreover, narrow fusion windows for the
synthesis are expected also for hot-fusion reactions. On the basis
of these considerations we expect the fusion window for the
synthesis of SHE 118 in the collision $^{48}$Ca on $^{252}$Cf
around 206 MeV.
\begin{figure*}
% Use the relevant command for your figure-insertion program
% to insert the figure file.
% For example, with the option graphics use
\begin{center}
\resizebox{13.7cm}{!}{%
 \includegraphics{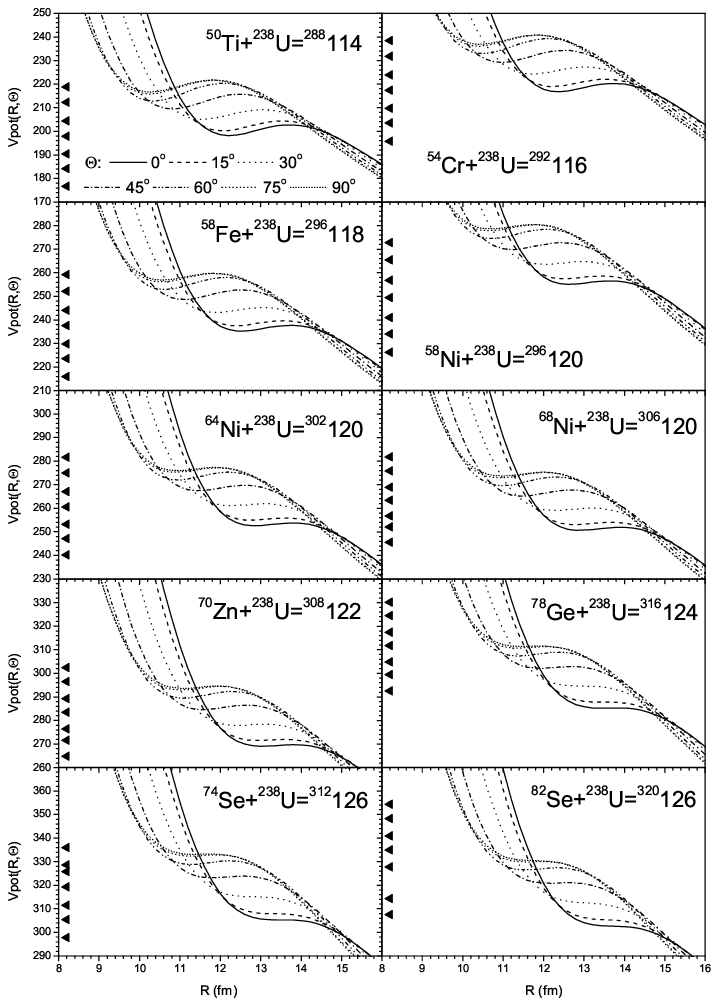}
}
% If not, use
%\vspace{0cm} % Give the correct figure height in cm
\end{center}
\caption{The SMPs for the systems $^{50}$Ti, $^{54}$Cr, $^{58}$Fe,
$^{58}$Ni, $^{64}$Ni, $^{68}$Ni, $^{70}$Zn, $^{78}$Ge, $^{74}$Se,
$^{82}$Se + $^{238}$U evaluated with the SkM$^*$ Skyrme force. The
SMPs are evaluated for different angular orientations of
$^{238}$U. The notations are the same as in Fig. 4. }
\label{fig:6} % Give a unique label
\end{figure*}

Since the $\alpha$-decay chain of the superheavy isotopes, formed
in the $^{48}$Ca induced reactions, do not reach the area of known
isotopes, it is difficult to decide on the isotope formed in the
synthesis. Lighter Ca isotopes ($^{40,42}$Ca) would lead to
isotopes formed earlier in cold-fusion reactions, if the number of
evaporated neutrons in $^{48}$Ca and $^{40,42}$Ca induced
reactions are the same. Although the pockets have the same or even
slightly larger depths, the excitation energies are considerably
larger for $^{40,42}$Ca as compared to the $^{48}$Ca induced
reactions. Therefore SHE formation is probably too much suppressed
for reactions with $^{40,42}$Ca projectiles.

\subsubsection{Systems with $^{238}$U and $^{252}$Cf targets}

As discussed in the preceding section 3.2.1, hot-fusion systems
with uranium and transuranium targets may be good candidates for
the synthesis of SHEs heavier than those reached with $^{208}$Pb
targets. Like for $^{208}$Pb we study in this section a series of
projectiles with increasing mass for $^{238}$U and $^{252}$Cf
targets. $^{238}$U has been chosen as one of the most convenient
targets with respect to radioactivity, availability and target
properties. Moreover it is one of the largest-bound isotopes in
this mass region, and hence is in this respect similar to
$^{208}$Pb yielding relatively small compound-nucleus excitation
energies. $^{252}$Cf has been depicted as the heaviest target
available \cite{sl}.

\begin{figure*}
% Use the relevant command for your figure-insertion program
% to insert the figure file.
% For example, with the option graphics use
\begin{center}
\resizebox{13.7cm}{!}{%
 \includegraphics{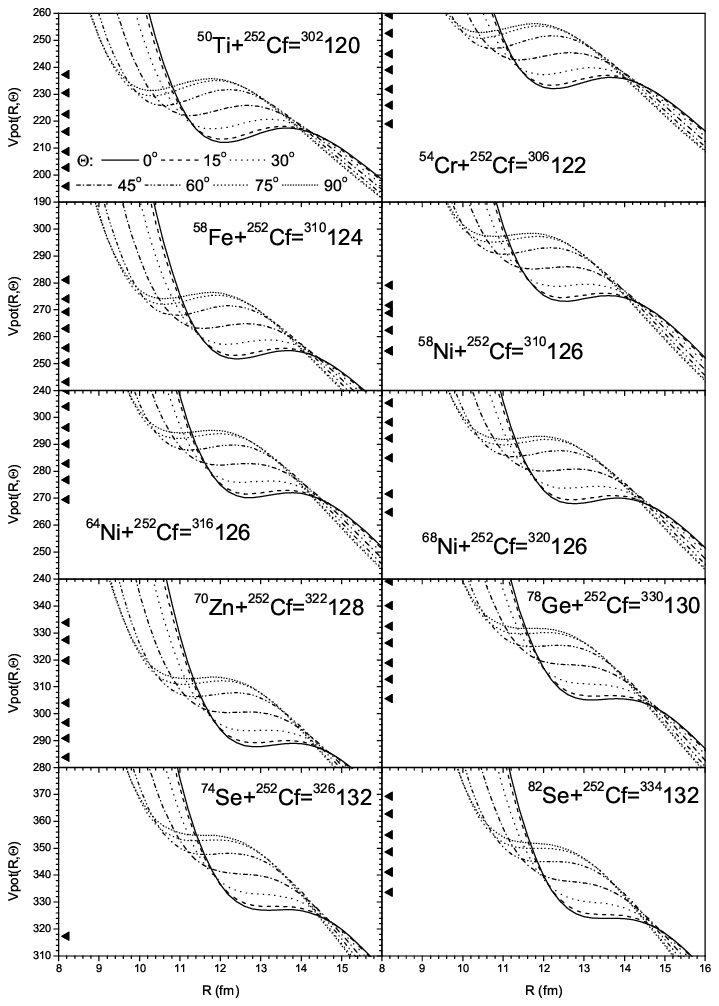}
}
% If not, use
%\vspace{0cm} % Give the correct figure height in cm
\end{center}
\caption{The SMPs for the systems $^{50}$Ti, $^{54}$Cr, $^{58}$Fe,
$^{58}$Ni, $^{64}$Ni, $^{68}$Ni, $^{70}$Zn, $^{78}$Ge, $^{74}$Se,
$^{82}$Se + $^{252}$Cf evaluated with the SkM$^*$ Skyrme force.
The SMPs are evaluated for different angular orientation of
$^{252}$Cf. The notations are the same as in Fig. 4.}
\label{fig:7} % Give a unique label
\end{figure*}

The entrance-channel potentials are shown in Figs. 6 and 7 for
$^{238}$U and $^{252}$Cf targets, respectively. The dependence on
the orientation of the deformed nuclei is similar to that of the
deformed targets with $^{48}$Ca and $^{40,42}$Ca in Figs. 4 and 5.
Again, for the side orientation ($\Theta = 90^\circ$) the barrier
and pocket shapes are considerably more compact than for
$^{208}$Pb and by about 25 MeV higher in energy. The general
decrease of the depth and width of the pockets with increasing
charge of the projectile is similar to that for the $^{208}$Pb
target in Fig. 3.

Some interesting points of Fig. 6 for $^{238}$U are summarized as
follows.
\begin{itemize}

\item[--] The potential pocket vanishes for the $^{78}$Ge
projectiles leading to the SHE $^{316}$124, whereas this situation
happens already for $^{96}$Zr on $^{208}$Pb forming SHE
$^{304}122$.

\item[--] The potential pocket for $^{58}$Fe+$^{238}$U is also
somewhat larger than the one for $^{86}$Kr+$^{208}$Pb, both
leading to SHE 118. However, higher excitation energies are
encountered for bombarding energies at the barrier of
$^{58}$Fe+$^{238}$U.

\item[--] Comparing the entrance-channel potentials for different
nickel isotopes on $^{238}$U, we notice no difference in the
shape; but the height of the barrier, measured with respect to the
compound-nucleus ground-state energy, reduces considerably with
increasing neutron number. This effect has been recognized already
by comparing $^{40,42}$Ca with $^{48}$Ca in Figs. 4 and 5.
Therefore the use of more neutron-rich projectiles may be in
general favorable for SHE formation, because the compound nucleus
is formed with less excitation energy.

\end{itemize}

For the series of projectiles on $^{252}$Cf we recognize from Fig.
7 that the potential pockets are considerably deeper as compared
to those with $^{238}$U for the same compound system, and even
somewhat deeper for the same projectile. The pocket for the side
orientation vanishes for $^{78}$Ge+$^{252}$Cf $=^{330}$130 while
this happens already for the compound system $^{316}$124 if
$^{252}$Cf is replaced by $^{238}$U. In addition, also the
compound-nucleus excitation energies are lower for the $^{252}$Cf
target. From all these features we expect larger cross-sections
for the synthesis of the SHEs with $^{252}$Cf than with $^{238}$U
targets. If we consider the pocket of the side orientation to be
decisive, the limit for fusion is reached around
$^{70}$Zn+$^{252}$Cf=$^{322}$128.

\subsection{Warm-fusion systems}

Recently $^{233,234}$Cm and Fm have been synthesized in the
collisions $^{40}$Ar,$^{50}$Ti+$^{198}$Pt \cite{hofmann_pt}. The
cross-section for $^{50}$Ti+\\$^{198}$Pt is comparable with the
one for the cold-fusion reaction $^{40}$Ar+$^{208}$Pb. $^{198}$Pt
is oblate with deformation constant $\beta_2=-0.10$ \cite{pt198}
and offers a qualitatively new entrance channel for the synthesis
of SHEs.

In Fig. 8 we present the entrance-channel potentials for $^{40}$Ar
till $^{100}$Mo + $^{198}$Pt. In contradistinction to the prolate
targets $^{238}$U, $^{244}$Pu, $^{248}$Cm and $^{252}$Cf the
dependence of the barrier on the orientation is opposite with the
lowest barrier for $\Theta=90^\circ$ (side position) and the
highest barrier for $\Theta=0^\circ$ (tip position), the
difference being about 10 MeV. The difference between the highest
and lowest barriers for $^{198}$Pt is smaller than the one for
uranium and transuranium cases due to the smaller $|\beta_2|$.
Furthermore, the pockets are considerably deeper for the tip
position than for the side position.

The larger tip-position pockets for the oblate shape as compared
to the spherical and prolate shapes is essentially due to the
small curvature of its surface at the tip position. Indeed, the
interaction potential in the proximity approach
\cite{prox2000,prox77} around touching is proportional to the
reduced radii of the surfaces, i.e.
$$V_N \propto \frac{R_1 R_2}{R_1+R_2} $$
with $R_1, R_2$ the radii of the near touching surfaces. If
nucleus 2 is deformed, the curvature $C_2=1/R_2$ is replaced by
the mean curvature $\overline{C}_2=1/\overline{R}_2$. We consider
a spheroid with half axes $a=b=R_2(1-\delta)$ and
$c=R_2(1+2\delta)$. For the same mass and absolute value of the
deformation parameter $\delta$ we have
$$\overline{R}_2^{\rm ps}=R_2(1+2\delta)$$ with $\delta > 0$ for the side
position of the prolate nucleus and
$$\overline{R}_2^{\rm ot}=R_2(1+4|\delta|)$$
with $\delta < 0$ for the tip position of the oblate nucleus.
Therefore as compared to spherical nuclei, the pockets, which are
caused by the nuclear attraction between the surfaces of the
nuclei, are deeper for the side position of prolate nuclei and
even more for the tip position of oblate nuclei. The Coulomb
interaction favors the depths of the pockets for the oblate tip
position as well as for the prolate side position. %Outside the
%range of nuclear interaction the Coulomb interaction tend to
%rotate the oblate nucleus into the $\Theta =0^\circ$ position (tip
%position) and increase the (oblate) deformation \cite{denisov1}.

\begin{figure*}
% Use the relevant command for your figure-insertion program
% to insert the figure file.
% For example, with the option graphics use
\begin{center}
\hspace{-1cm}\resizebox{17cm}{!}{%
 \includegraphics{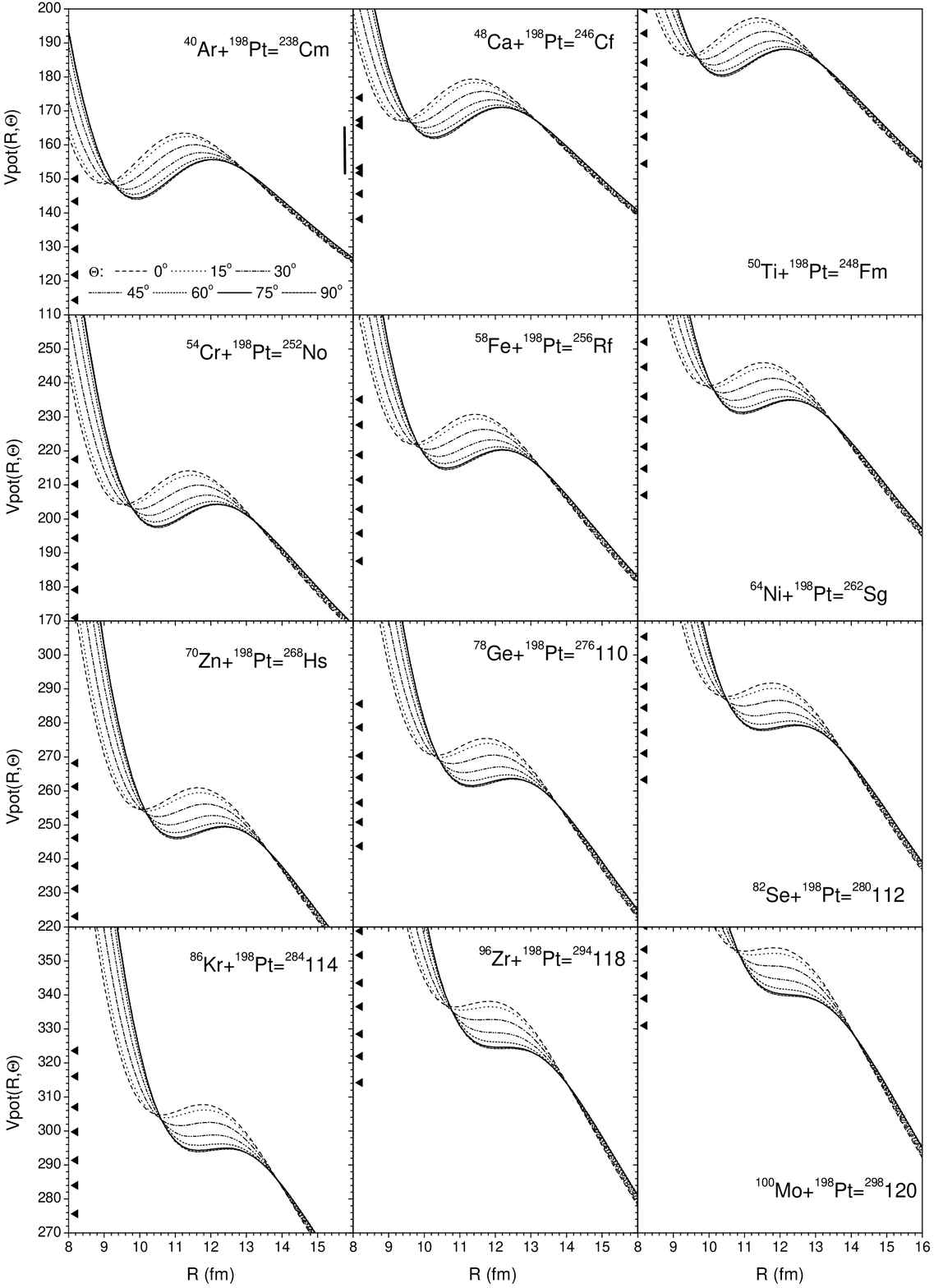}
}
% If not, use
%\vspace{0cm} % Give the correct figure height in cm
\end{center}
\caption{The SMPs for the systems $^{40}$Ar, $^{48}$Ca, $^{50}$Ti,
$^{54}$Cr, $^{58}$Fe, $^{64}$Ni, $^{70}$Zn, $^{78}$Ge, $^{82}$Se,
$^{86}$Kr, $^{96}$Zr, $^{100}$Mo + $^{198}$Pt evaluated with
SkM$^*$ Skyrme force. The SMPs are evaluated for different angular
orientation of $^{198}$Pt. The notations are the same as in Fig.
4.}
\label{fig:8} % Give a unique label
\end{figure*}

In general for the same compound nucleus, the barrier energies
measured with respect to the ground-state compound energy lie for
$^{198}$Pt in between the barriers for the cold-fusion target
$^{208}$Pb and the hot-fusion targets $^{238}$U, $^{244}$Pu,
$^{248}$Cm and $^{252}$Cf, and hence we refer to $^{198}$Pt as the
warm-fusion target. Here we imply that the SHE fusion window is
again located about 5 to 10 MeV below the SMP barriers of the most
compact pockets in the tip position.

\section{Conclusion}

As pointed out already in the introduction, the capture of the
colliding nuclei within the pocket of the interaction potential is
a decisive first step in the fusion process. For the heavy systems
under consideration these pockets are rather shallow and vanish
for too large systems. The population of long-living quasi-bound
states in the pocket is expected to be limited in bombarding
energy from below by the barrier penetrability (including transfer
channels and polarization of the nuclei). For incident energies
above the barrier, reflection of the system by the large repulsive
core of the SMP potential prevents the capture in the pocket. The
resulting window in bombarding energy for capture should increase
with increasing depths of the potential pockets. These arguments
are consistent with the observed fusion window in the collision
energy which is typically 5 MeV wide and lies about (5 -- 10) MeV
below our SMP barrier. Note, however, that the observed fusion
window for SHE formation may differ from the capture window by the
additional limitation towards higher energies due to
$\Gamma_n/\Gamma_f \ll 1$.

There is a second limitation of the fusion window from below,
because the capture has to be at high enough energies, such that
the system can reach the compound-nucleus shape and fall into the
potential minimum by emitting one neutron. The experimentally
observed squeezing \cite{hm,hofmann} of the fusion window for
$^{64}$Ni + $^{208}$Pb and $^{70}$Zn + $^{208}$Pb may be ascribed
to this effect. From this point of view the synthesis of $^{86}$Kr
+ $^{208}$Pb should be suppressed, because the energies necessary
for the 1n fusion is above the capture window (cf. Fig. 2).

On the basis of these arguments we define the following rules for
the determination of the best candidates for the synthesis of
SHEs.

\begin{itemize}
\item[--] The SMP barrier should lie about 10 to 15 MeV above the
1n fusion threshold, but not above the 2n fusion threshold to
avoid the reduction of the fusion cross-section by an additional
factor $\Gamma_n/\Gamma_f$. This condition yields an optimum
fusion window (5 to 10 MeV below the barrier) for the formation of
a compound nucleus with excitation energy 5 MeV above the
1n-emission threshold.

\item[--] The pocket depth should be as large as possible, because
the deeper the pocket is, the larger the capture window becomes,
and hence the better is the chance of fusion.

\item[--] For the subsequent formation of a compound nucleus it is
best to have a most compact capture configuration.
\end{itemize}

We illustrate these rules for the synthesis of SHE 118 with hot-,
cold- and warm-fusion systems studied in section 3.

\begin{itemize}
\item[--] As mentioned already, the cold-fusion system
$^{86}$Kr+\\$^{208}$Pb (cf. Fig. 2) has its capture window below
the 1n-fusion channel, and hence is not expected to be a good
candidate.

\item[--] The symmetric system $^{144}$Ce+$^{150}$Nd (cf. Fig. 3)
has no pocket, and hence no capture window at all.

\item[--] The hot-fusion system $^{48}$Ca+$^{252}$Cf (cf. Fig. 4)
has nice capture properties, however needs to emit about 2 to 3
neutrons, which reduce the survival probability by several orders
due to factors $\Gamma_n/\Gamma_f \ll 1$.

\item[--] The hot-fusion system $^{40}$Ca+$^{252}$Cf (cf. Fig. 5)
has less attractive capture properties (as compared to the
$^{48}$Ca case) and needs to emit even 5 to 6 neutrons.

\item[--] The system $^{58}$Fe+$^{238}$U (cf. Fig. 6) has only a
tiny pocket and needs to emit about 3 to 4 neutrons.

\item[--] The warm-fusion system $^{96}$Zr+$^{198}$Pt has also a
tiny tip-positioned pocket, but needs to emit only one neutron.
\end{itemize}
From this we conclude that among the studied systems the most
attractive projectile-target combinations for the synthesis of SHE
118 are $^{48}$Ca+$^{252}$Cf at $E_{\rm coll}\approx 206$ MeV
(side collision) and $^{96}$Zr+$^{198}$Pt at $E_{\rm coll}\approx
330$ MeV (tip collision). While $^{48}$Ca+$^{252}$Cf is more
compact, $^{96}$Zr+$^{198}$Pt needs to emit only 1 neutron instead
of 2 to 3. It is hard to judge, which of these features are more
important for the synthesis of SHE 118. Generally for deformed
target nuclei, the experiments on the synthesis of SHEs may profit
from using polarized targets by aligning prolate targets for side
collisions and oblate targets for tip collisions.

One should be aware, that other effects not due to capture,
compactness and neutron emission may influence the final formation
of a SHE. In particular special features of the potential
landscape, e.g. with respect to decay by quasi-fission, are of
importance. We hope to obtain an improved guidance for experiments
from a scattering model which treats capture, collective evolution
and neutron decay on the same level and fully quantum-mechanically
\cite{nd}.

\section*{Acknowledgements}

The authors would like to thank P. Armbruster and S. Hofmann for
fruitful discussions, and P. Rozmej for his help in working with
the spherical Hartree-Fock-Bogoliubov code. One of us (V.Y.D.)
gratefully acknowledges support from GSI.

\end{document}